# Title: The Valence-Fluctuating Ground State of Plutonium

**Authors:** M. Janoschek[1]*, Pinaki Das[1]†, B. Chakrabarti[2], D. L. Abernathy[3], M. D. Lumsden[3], J. M. Lawrence[1], J. D. Thompson[1], G. H. Lander[4], J. N. Mitchell[1], S. Richmond[1], M. Ramos[1], F. Trouw[1], J.-X. Zhu[1], K. Haule[2], G. Kotliar[2], E. D. Bauer[1]

**Affiliations:**

[1]Los Alamos National Laboratory (LANL), Los Alamos, NM 87545, USA.

[2]Department of Physics and Astronomy and Center for Condensed Matter Theory, Rutgers University, Piscataway, NJ 08854-8019, USA.

[3]Quantum Condensed Matter Division, Oak Ridge National Laboratory, Oak Ridge, TN 37831-6475, USA

[4]European Commission, Joint Research Centre, Institute for Transuranium Elements, Postfach 2340, D-76125 Karlsruhe, Germany

*Correspondence to: Marc Janoschek (mjanoschek@lanl.gov)

†Current address: Ames Laboratory, US DOE, and Department of Physics and Astronomy, Iowa State University, Ames, Iowa 50011, USA

**Abstract**: A central issue in material science is to obtain understanding of the electronic correlations that control complex materials. Such electronic correlations frequently arise due to the competition of localized and itinerant electronic degrees of freedom. While the respective limits of well-localized or entirely itinerant ground states are well-understood, the intermediate regime that controls the functional properties of complex materials continues to challenge theoretical understanding. We have used neutron spectroscopy to investigate plutonium, which is a prototypical material at the brink between bonding and non-bonding configurations. Our study reveals that the ground state of plutonium is governed by valence fluctuations, that is, a quantum-mechanical superposition of localized and itinerant electronic configurations as recently predicted by dynamical mean field theory. Our results not only resolve the long-standing controversy between experiment and theory on plutonium's magnetism, but also suggest an improved understanding of the effects of such electronic dichotomy in complex materials.

## Introduction

Plutonium (Pu) is known for the instability of its nucleus, allowing it to undergo fission. The electronic cloud surrounding the Pu nucleus, however, is equally unstable due to the near degeneracy of multiple electronic configurations allowed by its special position in the periodic table. Pu belongs to the actinide series in which the 5*f*-electron shell is progressively filled. In the early part of this series stretching from Th to Np, the *5f* electrons are delocalized and thus contribute to bonding between neighboring atoms similar to the *5d* series. Atomic volumes decrease with increasing atomic number Z, reflecting the increased screening of the positive nuclear charge with each additional electron. In contrast, for much larger and heavier actinides (Am and beyond), the 5*f* electrons are well-localized and do not participate in bonding, and their atomic volumes decrease much slower with Z, as in the 4*f* lanthanide series. The 5*f* electrons of Pu—situated between Np and Am— exist in the abyss between these two opposing tendencies, making Pu the most electronically complex element in the periodic table, with intriguingly intricate properties for an allegedly simple elemental metal that have defied understanding since the1940s (*1, 2*).

Because of this complexity, Pu exhibits a record-high number of six allotropic phases with large volumetric changes between these phases of up to 25%, and mechanical properties ranging from brittle α-Pu to ductile δ-Pu (*3*). The radius of Pu atoms in the face centered cubic δ-phase is midway between that of Np and Am. δ-Pu exhibits a temperature-independent, Pauli-like magnetic susceptibility and a Sommerfeld coefficient of the specific heat that are an order of magnitude larger than in any other elemental metal (*4*) due to the strong electronic correlations that emerge from the delicate interplay of itinerant and localized electronic degrees of freedom (*5*). Even beyond Pu, it is recognized that the understanding of strong electronic correlations is a key issue for complex materials in general (*6*). However, the description of their electronic ground state continues to pose a significant challenge to theory, precisely because such materials exist in between the well-understood extremes of electron localization/delocalization (*7*).

The conundrum of Pu's electronic instability becomes most clear from the drastic disagreement between conventional electronic structure theory and experiments (*4*). The large, temperature-independent magnetic susceptibility of Pu implies the absence of a net static magnetic moment expected if the 5*f* electrons were localized and is consistent with muon-spin rotation experiments that set an upper limit for a static or even slowly fluctuating (on a time scale of μsec and longer) moment of ≤$10^{-3}$ $\mu_B$/Pu (*8*). In contrast, conventional theories that succeed in correctly accounting for the structural and volumetric changes between the various phases of Pu predict static magnetic moments varying from 0.25 to 5 $\mu_B$/Pu [see Ref. (*4*) and references therein].

Resolving the ground state of the δ-phase of Pu, which shows notoriously complex behavior despite its high-symmetry *fcc* crystal structure with only a single element, presents an excellent opportunity to isolate the effects of strong electronic correlations and make progress on their understanding. A promising solution was recently proposed by a dynamical mean field theory (DMFT) calculation of Pu's electronic structure (*9*), and consists of modelling the ground state as a quantum mechanical admixture of localized and itinerant electronic configurations. The question whether the ground state of δ-Pu is indeed a true quantum-mechanical superposition may only be answered via observation of the associated virtual valence (charge) fluctuations among the distinct 5$f^4$ ($Pu^{4+}$), 5$f^5$ ($Pu^{3+}$), and 5$f^6$ ($Pu^{2+}$) electronic configurations. Here we reveal these valence fluctuations via inelastic neutron scattering, thus resolving the long-standing controversy about its electronic complexity and "missing magnetism".

In the presence of strong electron-electron correlations, the *f* electron wave function is typically well localized giving rise to magnetic moments as shown in Fig. 1A. Because the aforementioned theories reproduce the correct experimental densities for Pu via 5*f* electron localization, they predict a magnetically ordered state. However, below a Kondo temperature $T_K$, itinerant conduction electrons tend to align their spins antiparallel with respect to the 5*f* magnetic moment that in turn becomes compensated (Fig. 1B) (*10*). Through this dynamic interaction in the spin degrees of freedom, the *f* electron becomes hybridized with the conduction electrons, effectively leading to its delocalization into the Fermi conduction sea where it forms a heavy quasiparticle. This results in a strongly modified electronic density of states that then includes the electronic *f* level as a quasiparticle resonance with a width of $k_B T_K$ ($k_B$ is the Boltzmann constant) at the Fermi level $E_F$ (Fig. 1C).

Our DMFT calculations, which treat the effects from all important energy scales, notably the Kondo interaction, atomic multiplet effects and crystal field splitting equally, and the electronic band structure in a self-consistent way (*11*), demonstrate that the quantum-mechanical mixing of the different valence configurations dominates the physics of Pu. Notably, the 5*f* electrons are continuously hopping into and out of the Fermi sea via the quasiparticle resonance resulting in virtual valence fluctuations. We note that each of the quantum-mechanically admixed 5*f* states hybridizes with the conduction electrons. For example the main resonance peak at $E_F$ (see Fig. 1C) is predominantly due to fluctuations between a $f^5$ state with total spin *J* = 5/2 and the $f^6$ state with total spin *J* = 0; whereas, the two lower resonances at approximately 0.6 and 0.9 eV are due to fluctuations between higher total spin states of the $f^5/f^6$ configurations (*9*).

Core-hole photoemission spectroscopy (*12*) and resonant x-ray emission spectroscopy (RXES) measurements (*13*), both of which probe the valence configuration on a very short time scale ($\tau \approx 1$ fs), find a multi-valence ground state in δ-Pu suggested by DMFT calculations (*9*), with reasonable agreement for the occupation of the $5f^4$, $5f^5$, and $5f^6$ states (Table 1). Because these measurements only allow an essentially instantaneous snapshot of the electronic configuration, the corresponding virtual valence fluctuations remained hidden. In contrast, neutron spectroscopy is sensitive to the expected time scale of approximately 0.01 ps upon which spin fluctuations develop from virtual interconfigurational excitations of the Pu ion from the magnetic $f^5$ ($J= 5/2$) to the non-magnetic $f^6$ ($J= 0$) state.

The physics of spin fluctuations driven via valence fluctuations is captured by the Anderson impurity model (AIM) that describes the interaction of a magnetic impurity with a "bath" of conduction electrons (*14*), and of which the Kondo impurity problem is a special case (*15*). We note that while δ-Pu is actually a dense Kondo lattice in which interactions between a periodic array of Kondo 'impurities' lead to lattice effects, previous work on Kondo lattice compounds has shown that generally 80-90% of the magnetic-fluctuation spectrum is still correctly described by the AIM (*16*). For temperatures $T < T_K$ the 5$f$ spin dynamics of a Kondo impurity are those of a localized, damped oscillator with a characteristic spin-fluctuation energy $E_{sf} = k_B T_K$ resulting in dynamic magnetic susceptibility of the form (*16-19*)

$$\chi''(\omega) \propto \frac{\chi(T)(\hbar\omega)\Gamma}{(\hbar\omega - E_{sf})^2 + \Gamma^2}. \qquad (1)$$

Here $\hbar\omega$ is the energy transferred to the material with $h = 2\pi\hbar$ being the Planck constant, and $\chi(T)$ describes the temperature dependence of the susceptibility. $\Gamma$ is inversely proportional to the lifetime $\tau$ of the fluctuations via $\tau = \hbar/2\Gamma$. Earlier DMFT results yield $T_K \approx 800$ K (*9*), and we therefore would expect to observe a spin resonance characterized by $E_{sf} = 66$ meV.

**Results**

Figure 1D shows the dynamic magnetic susceptibility $\chi''(\omega)$ of δ-Pu derived from our experiment with an incident neutron energy $E_i = 500$ meV and at room temperature ($T = 293$ K) (*11*). A clear resonance-like feature is characterized by a spin fluctuation energy $E_{sf} = 84(1)$ meV in good agreement with the earlier DMFT results (*9*). This corresponds to a Kondo temperature $T_K = 975$ K. There is another feature with a higher spin-fluctuation energy of approximately 146 meV. It arises due to Kondo lattice effects that are accounted for in our state-of-the-art, self-consistent DMFT calculations (*11*), as illustrated by the calculated $\chi''(\omega)$ in Fig. 1D (dashed-blue line). We note that the position of the main peak, the linear fall off at low frequencies and the broad distribution of the spectral weight extending to high energies, are all very robust features of both the DMFT calculations and the experimental data. Notably, theory and experiment are in quantitative agreement if one takes into account their respective uncertainties as detailed in (*11*). Figure 1E shows the momentum ($Q$) and energy ($\hbar\omega$) transfer dependence of the observed magnetic scattering that is given by $F^2(Q)\chi''(\omega)$, where $F(Q)$ is the magnetic form factor for Pu. To confirm that the observed dynamic susceptibility is not an experimental artifact, we performed a second experiment with an incident energy $E_i = 250$ meV. Apart from differences in the experimental resolution, the same $\chi''(\omega)$ is obtained (*11*). We note that crystal field excitations or intermultiplet transitions would, in principle, lead to similar forms of $\chi''(\omega)$ but can be ruled out from arguments described in (*11*). We have also computed $F^2(Q)\chi''(\omega)$ which is in excellent agreement with our measurements (Fig.1F).

Using Eq. (1) we fit the dynamic susceptibility (red solid line in Fig. 1D) and extract the lifetime of the intertwined valence and spin fluctuations. We note that each of the two maxima is hereby fitted separately via the Lorentzian peak shape described by Eq. (1). The lifetime of both features agree within the error bars, where for the main feature with $E_{sf} = 84$ meV, we obtain be $\tau = 0.015(4)$ ps ($\Gamma = 28.4(9)$ meV). This explains why the magnetic fluctuations were not previously observed by the muon spin rotation measurements that only probe longer time scales down to 1 ps. Further, by integrating the observed intensities appropriately (*11*), we determine the size of the fluctuating moment as $\mu = 0.6(2)$ $\mu_B$. Using the effective moment of the $5f^4$, $5f^5$, and $5f^6$ states, as well as their occupation probabilities determined by x-ray spectroscopy (cf. Table 1), the fluctuating moment should be $\mu = 0.8$ $\mu_B$. We note that this intermediate coupling free-ion value only gives an upper limit because it neglects the possible influence of crystal fields and conduction electrons that may account for the difference observed here. Similarly, the DMFT calculation yields a fluctuating moment of $\mu = 0.82$ $\mu_B$ [see (11), particularly Eq. S16 and Fig. S5] in good agreement with experiment.

Figure 2 plots $F(Q)$ obtained by energy-integrating both data sets in the range $\pm 10$ meV around $E_{sf}$. For comparison, we show the tabulated magnetic form factors for $5f^4$ and $5f^5$ configurations (*20*), assuming an intermediate coupling scheme (*21*). The $5f^6$ ground state is non-magnetic and its contribution to the δ-Pu magnetic form factor is therefore negligible (*11*). A $5f^4$ magnetic form factor (dashed red line) cannot explain our data; whereas, a pure $5f^5$ magnetic form factor (solid red line) describes the data very well. However, both the experimental and theoretical average occupancy of the $5f^5$ configuration is far less than unity (see Table 1). Taking this into account, a weighted sum according to the $5f$ state occupation derived from RXES in Table 1 (dash-dotted red line) also reproduces the observed form factor. Finally, the form factor calculated via DMFT (blue solid line) that implicitly includes the $5f$ occupations is also in good agreement with the experiment, and supports the conclusion that the ground state of δ-Pu is indeed a multi-valence state.

Using the sum-rule for the dynamic magnetic susceptibility $\chi'(0) = \frac{1}{\pi}\int d\omega \frac{1}{\omega}\chi''(\omega)$ (*11*) we demonstrate that the magnetic properties of δ-Pu are consistently described by a valence-fluctuating ground state. Here $\chi'(0)$ is the static magnetic susceptibility, which from magnetic susceptibility measurements is known to be almost temperature-independent with a value of $\chi_{Bulk} = 5.3 \cdot 10^{-4}$ cm$^3$/mol at room temperature (*22*). Evaluating the sum rule for the dynamic susceptibility, our neutron scattering experiments yield $\chi^{f4+f5}_{Neutron} = 0.8(2) \cdot 10^{-4}$ cm$^3$/mol, where we note that neutrons are insensitive to the temperature-independent Van-Vleck susceptibility $\chi^{f6}_{VV}$ of the non-magnetic $5f^6$ state. However, we estimate $\chi^{f6}_{VV} = 3.1(15) \cdot 10^{-4}$ cm$^3$/mol from the magnetic susceptibility of the pure $5f^6$ state of Am, $\chi^{f6}_{Am} = 8.3 \cdot 10^{-4}$ cm$^3$/mol (*23*), which was scaled by the fractional $5f^6$ occupation by x-ray spectroscopy (Table 1). As demonstrated in Table 2, the sum of both contributions reproduces the measured static magnetic susceptibility of δ-Pu from Ref. (*22*) reasonably well. In this comparison we have not included the small temperature independent Pauli susceptibility of conduction electrons, which likely accounts for the small difference.

**Discussion**

In summary, the combination of our neutron spectroscopy and DMFT results unambiguously establishes that the magnetism in δ-Pu is not "missing", but dynamic, and is driven by virtual valence fluctuations.

Our measurements provide a straightforward interpretation of the microscopic origin of the large, Pauli-like magnetic susceptibility of δ-Pu and associated Sommerfeld coefficient. Several properties of δ-Pu have been successfully reproduced by phenomenological, so-called "two-level" models featuring ground states formed from a fixed admixture of two states (*24*). The experiments presented here, in combination with earlier core-hole spectroscopy (*12*) and RXES measurements (*13*) now define these "two levels" for the first time. Furthermore, because the various valence configurations imply distinct sizes of the Pu ion, the valence-fluctuating ground state of Pu also provides a natural explanation for its complex structural properties and in particular the large sensitivity of its volume to small changes in temperature, pressure or doping. As this work has shown, DMFT (*9, 11*) has reached a level of sophistication and control that it now can anticipate the ground state and related properties of a material as complex as δ-Pu, and is poised to be a useful predictive tool for the design and understanding of complex, functional materials that frequently are characterized by similar electronic dichotomies.

**Materials and Methods**

*Neutron Spectroscopy:* We note that previous neutron experiments carried out on δ-Pu by Trouw *et al.*(*25*) reported a resonance-like feature at approximately 90 meV in agreement with our study presented here. However, because of the strong neutron absorption of Pu and its frequent contamination with hydrogen that leads to strong spurious signals in the region of interest, as well as special double-wall sample containers that are required for safety reasons and lead to increased background signals, neutron experiments on Pu are challenging, and it remained unclear whether this feature stemmed from spin fluctuations. We have designed and carried out an experiment and analysis that overcomes all of these issues as described in the following.

To avoid the high neutron absorption cross-section of most Pu isotopes (see Table S1 in Ref. (*11*)) a δ-Pu sample with a total mass of 21.77 g (stabilized *fcc* structure with 3.5% atomic Ga, lattice parameter *a* = 4.608(1) Å, density $\rho$ = 15.81 g/cm$^3$) was prepared from 92.6% isotopically pure $^{242}$Pu (with less than 0.6% Pu-241), which is the least absorbing Pu isotope by more than an order of magnitude (see Table S1 S1 in Ref. (*11*)). The exact isotopics of the measured samples are given in Table S1 and results in a 1/e absorption length of 6 mm for thermal neutrons, thus allowing for a reasonable sized sample volume. In addition, the purity of this sample has been improved over previous samples used for neutron scattering (*4, 25, 26*) in a crucial way by removing hydrogen.

In order to remove hydrogen from the δ-Pu sample, it was placed in a Sieverts type apparatus and heated in vacuum for 72 hours at 450°C at which time the equilibrium hydrogen partial pressure indicated a hydrogen content of not more than 0.01 at. % based on Sieverts law (*27*) and it was subsequently cooled to room temperature. Following homogenization the rod was axially bisected and the samples were sent for metallography, density measurement and non-destructive assay of isotopic composition. Following these measurements and prior to final packaging the samples were once again vacuum-homogenized at 450°C for an additional 72 hours and hydrogen content was confirmed to be less than 0.01 at. %. The absence of sharp features in any of the spectra recorded throughout this experiment demonstrates that the hydrogen removal was successful. We note that the sample stayed in the δ stability regime during the entire procedure and never went above 450 °C.

The neutron spectroscopy measurements were performed at the instrument ARCS (*28*) at the Spallation Neutron Source operated at Oak Ridge National Laboratory (ORNL). A preliminary study was carried out at the Lujan Center at Los Alamos National Laboratory (LANL) using the PHAROS spectrometer. For safety and to avoid contamination the sample was sealed in a double-wall-Al can with indium seals using both screws and Stycast 2850FT epoxy, which contained an atmosphere of $^4$He exchange gas. The sample is a polycrystalline rod of approximately 6 mm diameter and 51 mm length and was cut into two half-cylinders that were mounted side by side on the Al plate (thickness less than 0.5 mm) attached in the center of the inner Al can in order to make use of the entire beam cross-section of ARCS. The flat side of the two half cylinders was perpendicular to the incoming beam.

The ARCS instrument is a direct geometry neutron time-of-flight chopper spectrometer. The instrument was used with two different incident energies $E_i$, namely 250 meV and 500 meV, resulting in energy resolutions at the elastic line of $\Delta E$ = 15 and 39 meV (FWHM, determined from vanadium standard measurements), respectively. We note that the inelastic resolution on the neutron energy loss side of the inelastic neutron scattering spectra is slightly improved, but as demonstrated in Ref. (*28*) this effect is only about 20% at the energy transfers of interest here (≈ 90 meV). To reduce the background from the double-wall-Al can while maintaining the scattered intensity of the sample the computer controlled aperture upstream of the sample position was set to 10 mm width and 50 mm height. In order to isolate the spin fluctuations a good estimate of the various background contributions, such as from the sample can, the phonon part of the inelastic spectrum, as well as multiple scattering is required. For this experiment this was solved by measuring an isostructural, non-magnetic analogue, as has been done similarly for other compounds such as CePd$_3$ (*16*) and Ce$_{1-x}$Th$_x$ (*29*), where La analogues have been used. Here we have used non-magnetic Th with a total mass of 20.371 g that was arc-melted into a similar shape of two half-cylinders mounted in an identical container that was measured with the same incident energies and identical statistics. In addition, we have measured an identical empty double-Al can to determine the background contribution produced solely from the container. As we show in detail in the supplementary materials (*11*), the low momentum transfer neutron scattering data that contains the details about the dynamic magnetic susceptibility presented in Fig. 1, is independent of the details of the used background subtraction, in turn, demonstrating the robustness of the presented results. All presented in this manuscript were carried out at room temperature.

*Dynamical Mean Field Theory (DMFT)*: The theoretical method for computing the magnetic response of correlated solids is based on Dynamical Mean Field Theory in combination with Density Functional Theory (DMFT+DFT) (*30*). We use the implementation of this method in its charge self-consistent and all electron methodology, as developed in Ref. (*31*). The DFT part is based on Wien2k package (*32*). In the DMFT+DFT method, the strong correlations on the Pu ion are described by the frequency and space dependent potential, called self-energy $\Sigma(\mathbf{r},\mathbf{r}',\omega)$, which is added to the DFT Kohn-Sham Hamiltonian to describe the entanglement of the Pu atomic states with the itinerant *spd* electrons as well as the neighboring Pu atoms. The self-energy contains all Feynman diagrams local to the Pu ion, and is defined through the quantum mechanical embedding in real space by:

$$\Sigma(\mathbf{r},\mathbf{r}',\omega) = \sum_{l=3,mm'} Y_{l,m}(\hat{\mathbf{r}}) R_l(\mathbf{r}) \Sigma_{lm,lm'}(\omega) R_l(\mathbf{r}') Y^*_{l,m'}(\hat{\mathbf{r}}'), \qquad (2)$$

where $R_l(\mathbf{r})$ is the radial part of the solution of the Dirac equation inside the Pu muffin-tin sphere (using Kohn-Sham like static self-consistent potential), linearized at the Fermi level. The components of the self-energy $\Sigma_{lm,lm'}$ are obtained by the solution of an auxiliary quantum impurity model, in which the impurity Green's function $G_{\text{impurity}}(\omega)$ and impurity self-energy $\Sigma_{\text{impurity}}(\omega)$ must match the corresponding quantities in the solid, i.e., $\Sigma_{lm,lm'}$ and $G_{lm,lm'}$, where the latter is given by

$$G_{lm,lm'}(\omega) = \left\langle Y_{lm} R_l \,|\, (\omega + \mu + \nabla^2 - V_{KS}(\mathbf{r}) - \Sigma(\mathbf{r},\mathbf{r}',\omega))^{-1} \,|\, R_l Y_{lm'} \right\rangle. \qquad (3)$$

The impurity model is solved by the numerically exact continuous time quantum Monte Carlo (CTQMC) method, as implemented in Ref. (*33*). Calculations are fully self-consistent in charge density, chemical potential and impurity levels, the lattice and impurity Green's functions, hybridizations, and self-energies. The partially screened Coulomb repulsion on Pu atom is $U = 4.5$ eV (*9, 34*) and Hund's coupling $J = 0.512$ eV. The simulations are performed at T = 232K, and lattice constant 4.61 Å, which corresponds to the experimentally determined fcc structure of δ-Pu.

On the order of 500 DFT and 30 DMFT cycles are required for self-consistency using the highly parallel leadership supercomputing resources of Titan. Of the order of 10 million core hours were used for high quality runs, which can be analytically continued to real frequencies with high confidence.

The magnetic susceptibility is computed in CTQMC by directly sampling the spin-spin correlation function in imaginary time defined by

$$\chi_{zz}(i\omega) = \int_0^\beta e^{i\omega\tau} \langle M_z(\tau) M_z(0) \rangle \quad (4)$$

where $M_z = \mu_B(L_z + 2S_z)$ is the magnetization on Pu atom. The details of the algorithm are given in Ref. (*33*). The convergence of the magnetic susceptibility with the number of Monte Carlo moves is as fast as the convergence of Green's function, hence high quality Matsubara data can be obtained. The real frequency susceptibility is obtained by analytic continuation using maximum entropy as well as Pade methods.

**Acknowledgments:**

**General**: B. C., G. K. and K. H. acknowledge fruitful discussion with Mario Pezzoli. M. J. acknowledges useful discussions with Amir Murani. We further thank Mark Taylor and Eric Larson at LANL for help with the design and engineering analysis of the double-wall-Al can. We are indebted to the technical, scientific and safety staff at ORNL; without their countless hours of meticulous planning in the two years prior to the experiment, and the excellent support during the experiment, this work would have not been possible.

**Funding:** Work at Los Alamos National Laboratory (LANL) was performed under the auspices of the U. S. Department of Energy. M. J., P. D., and E. D. B were funded by the LANL Directed Research and Development program. J.D.T. was funded by the DOE, Office of Basic Energy Sciences (BES). LANL is operated by Los Alamos National Security for the National Nuclear Security Administration of DOE under contract DE-AC52-06NA25396.

Research conducted at Oak Ridge National Laboratory's (ORNL) Spallation Neutron Source was sponsored by the Scientific User Facilities Division, Office of Basic Energy Sciences, US Department of Energy.

This research used resources of the Oak Ridge Leadership Computing Facility at ORNL, which is supported by the Office of Science of the US Department of Energy under Contract No. DE-AC05-00OR22725.

K. H. and B. C. is supported by NSF DMR-1405303. GK is supported by BES-DOE Grant DE-FG02-99ER45761.

G. H. L. acknowledges support by the Seaborg Institute at LANL.

**Author contributions:** M. J., P. D., D. L. A., M. D. K. and F. T. carried out the neutron scattering experiments. M. J. and J. D. T. analyzed data with input from J. M. L. and G. H. L. E. D. B., J. N. M., S. R., and M. R. prepared the sample. B. C., J. X. Z., K. H. and G. K. carried out the LDA+DMFT calculations. M. J. and E. D. B. designed the study with significant input from J. M. L. and G. H. L. M. J. wrote the manuscript and all authors discussed the results and commented on the manuscript.

# Figures and Tables

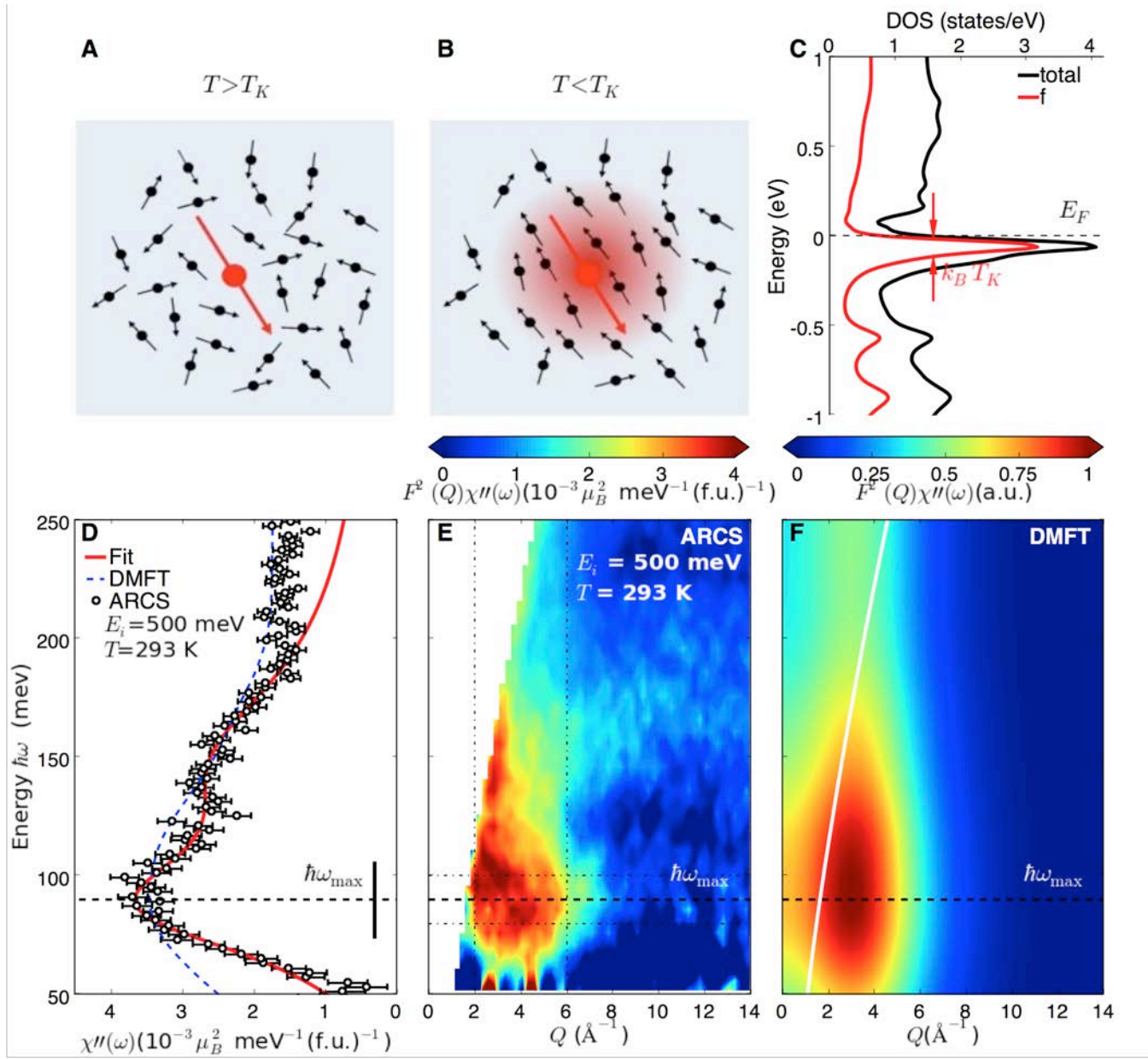

**Fig. 1**. **Visualization of the valence-fluctuating ground state of δ-plutonium (δ-Pu) by means of neutron spectroscopy.** (**A**) Above a characteristic Kondo temperature $T_K$ the $f$ electron wave function in $f$-electron materials such as Pu is typically well localized resulting in the formation of a magnetic moment (red). (**B**) For temperatures $T < T_K$ the conduction electrons (black) tend to align their spins antiparallel with respect to the magnetic moment that in turn becomes quenched, resulting in hybridization of the $f$ electron with the conduction electrons. As shown in (**C**) based on our Dynamical Mean Field Theory (DMFT) calculations for δ-Pu (see text), this leads to a strongly modified electronic density of states that then includes the electronic $f$ level as a "quasiparticle resonance" with a width of $k_B T_K$ ($k_B$ is the Boltzmann constant) at the Fermi level $E_F$. The DMFT calculation shows that the hybridization of $5f$ and conduction electrons drives a quantum-mechanical superposition of different valence configurations, where the $5f$ electrons are continuously hopping into and out of the Fermi sea via the quasiparticle resonance resulting in virtual valence fluctuations. Here we employ the spin fluctuations that arise from the repeated virtual ground state reconfiguration of the Pu ion from a magnetic (**A**) to a non-magnetic (**B**) state to visualize the valence fluctuations by measuring the dynamic magnetic susceptibility $\chi''(\omega)$ of δ-Pu by means of neutron spectroscopy. (**D**) $\chi''(\omega)$ obtained from our measurements carried out at room temperature ($T$ = 293 K) shows a maximum at the energy $\hbar\omega_{max} = \left(E_{sf}^2 + \Gamma^2\right)^{1/2}$ (black dashed line) that is determined by the characteristic spin-fluctuation energy $E_{sf} = k_B T_K$ and the lifetime $\tau$ of the fluctuations via $\tau = \hbar/2\Gamma$. The red solid line is a fit $\chi''(\omega)$ to Eq. (1) as described in the text. The broken blue line was calculated via DMFT. The vertical black bar represents the energy resolution of the experiment. (**E**) and (**F**) show the full energy ($\hbar\omega$) and momentum transfer ($Q$) dependence of the magnetic scattering as observed in our experiment and calculated by DMFT, respectively. The vertical and horizontal dash-dotted lines in (**E**) denote the integration ranges used for the energy and momentum transfer cuts shown in (**D**) and Fig. 2, respectively. The white solid line in (**F**) denotes the boundary beyond which no experimental data is available.

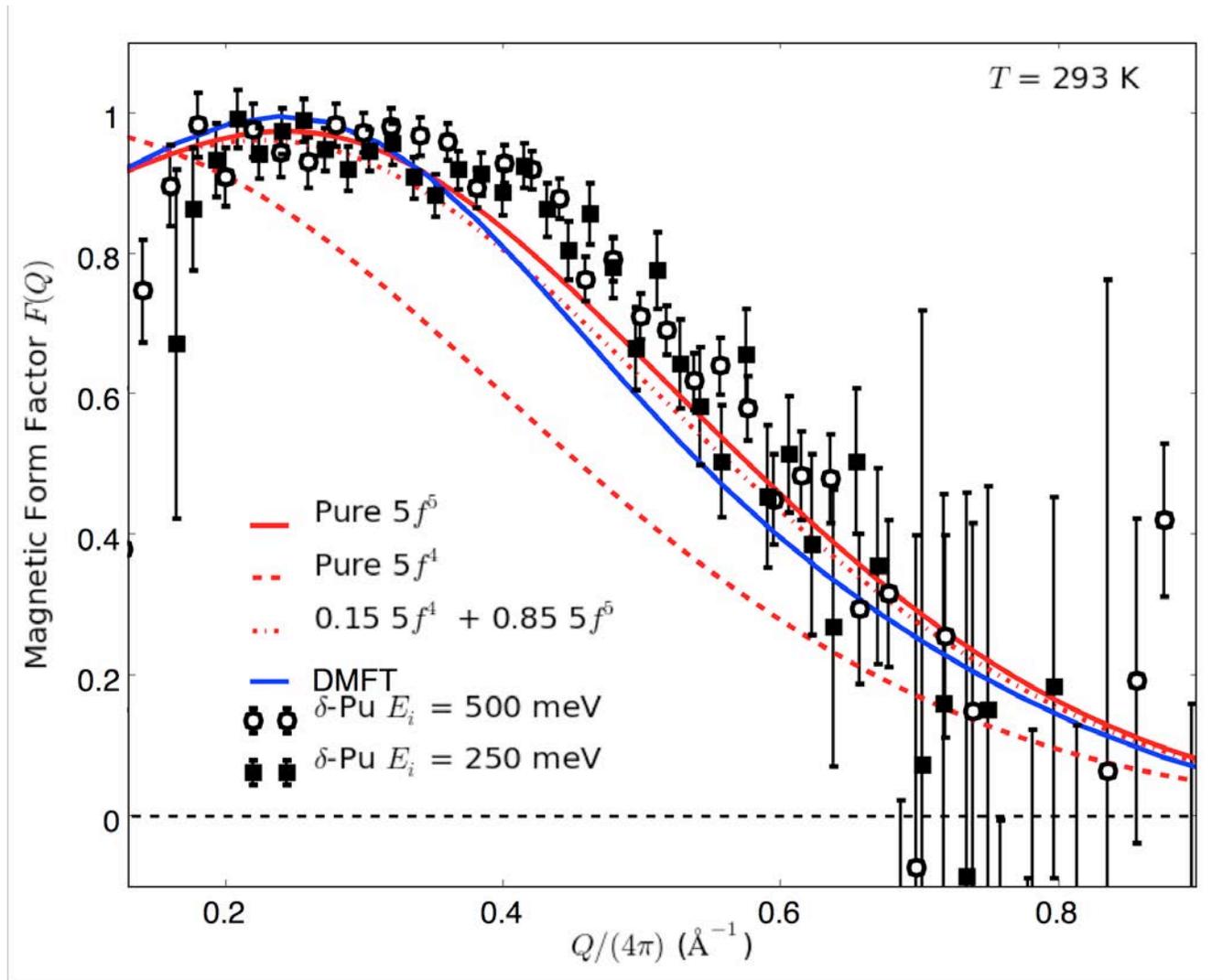

**Fig. 2. The magnetic form-factor for δ-plutonium (δ-Pu).** The black squares and open circles denote the magnetic form factor for δ-Pu as determined by our neutron spectroscopy experiment carried out at room temperature ($T$ = 293 K) and with incident neutron energies $E_i$ = 250 and 500 meV, respectively. The solid and dashed red lines are tabulated magnetic form factors for $5f^4$ and $5f^5$ electronic configuration in the intermediate coupling regime. The dashed-dotted red-line is a mix of both according to the $5f^4$ and $5f^5$ occupation as determined by resonant x-ray emission spectroscopy (see table 1). The solid blue line was calculated via dynamical mean field theory (*11*).

**Table 1. Average occupation of the 5*f* states in δ-Pu.** The occupation of the 5*f* states in δ-Pu is shown as calculated by (DMFT) and measured by resonant x-ray emission spectroscopy (RXES) (*13*) and core-hole photo-emission spectroscopy (CHPES) (*12*), respectively. We also list the corresponding effective moment $\mu_{eff}$ of the three 5*f* states based on the intermediate coupling scheme(*21*).

| δ-Pu 5*f* state | $f^4$ | $f^5$ | $f^6$ |
|---|---|---|---|
| **Occupation (DMFT) (%)** | 12 | 66 | 21 |
| **Occupation (RXES) (%)** | 8(2) | 46(3) | 46(3) |
| **Occupation (CHPES) (%)** | 6(1) | 66(7) | 28(3) |
| **Effective Moment $\mu_{eff}$ ($\mu_B$)** | 2.88 | 1.225 | 0 |

**Table 2. The various contributions to the magnetic susceptibility of δ-plutonium (δ-Pu).** $\chi_{Neutron}^{f4+f5}$ denotes magnetic susceptibility associated with the magnetic $5f^4$ and $5f^5$ states of δ-Pu at room temperature as determined by our neutron spectroscopy measurements. $\chi_{VV}^{f6}$ gives the temperature-independent Van-Vleck contribution $\chi_{VV}^{f6}$ of the non-magnetic $5f^6$ that is estimated from the published magnetic susceptibility of Am (see text). The error bar for $\chi_{VV}^{f6}$ was estimated from the various values of the occupation of the $5f^6$ state of δ-Pu determined by different experiments (Table 1). The sum of both reproduces the static magnetic susceptibility $\chi_{Bulk}$ of δ-Pu as measured by magnetic susceptibility reasonably well (*22*). All quantities are given in units of $10^{-4}$ cm$^3$/mol.

| $\chi_{Neutron}^{f4+f5}$ | $\chi_{VV}^{f6}$ | $\chi_{Neutron}^{f4+f5} + \chi_{VV}^{f6}$ | $\chi_{Bulk}$ |
|---|---|---|---|
| 0.8(3) | 3.1(15) | 3.9(15) | 5.3 |

# Title: Supplementary Materials for "The Valence-Fluctuating Ground State of Plutonium"

**Authors:** M. Janoschek[1]*, Pinaki Das[1]†, B. Chakrabarti[2], D. L. Abernathy[3], M. D. Lumsden[3], J. M. Lawrence[1], J. D. Thompson[1], G. H. Lander[4], J. N. Mitchell[1], S. Richmond[1], M. Ramos[1], F. Trouw[1], J.-X. Zhu[1], K. Haule[2], G. Kotliar[2], E. D. Bauer[1]

**Affiliations:**

[1]Los Alamos National Laboratory (LANL), Los Alamos, NM 87545, USA.

[2]Department of Physics and Astronomy and Center for Condensed Matter Theory, Rutgers University, Piscataway, NJ 08854-8019, USA.

[3]Quantum Condensed Matter Division, Oak Ridge National Laboratory, Oak Ridge, TN 37831-6475, USA

[4]European Commission, Joint Research Centre, Institute for Transuranium Elements, Postfach 2340, D-76125 Karlsruhe, Germany

*Correspondence to:  Marc Janoschek (mjanoschek@lanl.gov)

†Current address: Ames Laboratory, US DOE, and Department of Physics and Astronomy, Iowa State University, Ames, Iowa 50011, USA

## Supplementary Materials

**Neutron Diffraction Results:** For safety reasons the sample was loaded into the double-wall Al container at the Chemistry Metallurgy Research (CMR) facility at LANL, and could not be opened at the experiment at ORNL. Thus, to verify the integrity of the sample before starting the experiment a spectrum was recorded using an incident energy of $E_i$ = 40 meV. Integrating the result over all energy transfers a diffraction profile was obtained and fitted using Rietveld analysis implemented in the software FullProf (*35*). For the fits two different phases were used, namely one for the δ-Pu sample and one for the Al can as shown in Fig. S1. The profile could be fitted with reasonable quality considering the relatively poor angular resolution of ARCS that is not optimized for diffraction experiments, the fact that the sample was not a powder but an alloyed rod, and the relatively strong absorption of the sample, thus demonstrating that the sample survived the transport without any damage. The Pu lattice parameter extracted from the fit was $a$ = 4.608(1) Å, in good agreement the literature value for 3.5% Ga stabilized sample (*36*)

**Absorption Correction and Subtraction of Background:** As we will show in more detail below, the inelastic neutron spectra of the isostructural, non-magnetic analogue Th sample contains, apart from the magnetic contribution, the exactly same contributions as the δ-Pu spectrum, despite the slightly larger lattice parameter $a_{Th}$ = 5.08 Å. This is because for the high incident neutron energies employed here, this difference in the *inelastic* spectra is "smeared out" by the relaxed energy and momentum transfer resolution. This allows us to carefully subtract the various background contributions, such as the additional scattering from the double-wall-Al can, multiple scattering, and the phonon background, and obtain the pure magnetic contribution to the scattering from δ-Pu. To account correctly for the difference in absorption, the background generated from the double-Al can had to be subtracted from both spectra first considering the self-shielding for each sample, before the Th spectrum could be subtracted from the δ-Pu spectrum. We then obtain the magnetic contribution to the δ-Pu spectrum using the following equation:

$$I_{\text{Pu,mag}} = \left(I_{\text{Pu,mag}} - SSF_{\text{Pu}} I_{\text{can}}\right) - C\left(I_{\text{Th}} - SSF_{\text{Th}} I_{\text{can}}\right). \tag{S1}$$

Here $I_{\text{Pu}}$, $I_{\text{Th}}$, and $I_{\text{can}}$ denote the recorded spectra for the δ-Pu and Th samples, and the empty double-wall-Al can, respectively. $C$ is a correction factor that takes into account that δ-Pu and Th have different a coherent cross-section $\sigma$, density of scattering centers $mN_A/m_{\text{mol}}$ ($N_A$ is the Avogadro constant $m_{\text{mol}}$ is the molar mass, and $m$ is the mass), and is given by

$$C = \frac{\sigma_{\text{Pu}}}{\sigma_{\text{Th}}} \frac{m_{\text{Pu}}}{m_{\text{Th}}} \frac{m_{\text{mol,Th}}}{m_{\text{mol,Pu}}} \tag{S2}$$

Using the coherent scattering cross-section for our δ-Pu sample based on the measured isotopics given in Table S1, the coherent scattering cross-section for Th $\sigma_{Th}$ = 13.36 barn, the sample masses given in the previous section, and the molar weights $m_{\text{mol},Pu}$ = 241.886 g/mol (note this is the molar weight for the specific isotopic composition of the sample used here) and $m_{\text{mol},Th}$ = 232.038 g/mol, we obtain $C$ = 0.61.

The factors $SSF_{\text{Pu}}$ and $SSF_{\text{Th}}$ are self-shielding factors that take into account that if an absorbing sample is immersed in a neutron field, the interior of the sample will be exposed to a smaller neutron fluence rate than the exterior (*37, 38*). This has to be corrected for when the scattered intensity due to the double-wall-Al can is subtracted, but several analytical expressions for various sample shapes have been derived (*37, 38*). The sample used here is best approximated by a flat-plate geometry. In this case the self-shielding factor is given by

$$SSF = \frac{1 - e^{-x}}{x}, \tag{S3}$$

where $x = \Sigma_a t$. Here $\Sigma_a$ is the so-called macroscopic absorption cross-section in units of cm$^{-1}$ and $t$ is the thickness of the flat plate that we approximate to be 0.3 cm in the following. The macroscopic absorption cross-section can be calculated via

$$\Sigma_a = \rho N_a \sum_{i=1}^{N} \frac{w_i \sigma_i}{a_i}, \tag{S4}$$

where $\rho$ is the density of the sample, $N_a$ = 6.022 10$^{23}$ mol$^{-1}$ is the Avogadro constant, and $w_i$, $\sigma_i$ and $a_i$ are the relative weight, the absorption cross-section and atomic weight of the i*th* component of the sample, respectively. The absorption cross-sections for the different Pu isotopes tabulated in Table S1, as well as for Th are only valid for thermal neutrons with an energy $E$ = 25.3 meV (corresponding to a temperature $T$ = 293.6 K). The macroscopic absorption correction has to be adjusted for this using the following equation:

$$\Sigma_a(E) = \frac{2}{\sqrt{\pi}} \sqrt{\frac{25.3 \text{ meV}}{E[\text{meV}]}} \Sigma_a(25.3 \text{ meV}). \tag{S5}$$

A further small adjustment of $SSF$ is required because the sample also scatters neutrons, which may change the neutron's path length and hence the probability of absorption within the sample. This is achieved through the following empirical formula:

$$SSF = \frac{SSF_0}{1-\left(\Sigma_s/\Sigma_t\right)\left(1-SSF_0\right)}, \quad (S6)$$

with $\Sigma_t = \Sigma_s + \Sigma_a$. $\Sigma_s$ is the macroscopic scattering cross-section calculated analogues to $\Sigma_a$ (see Eq. (S4)). $SSF_0$ is calculated using Eq. (S3), however with $x = \Sigma_t t$. As shown in Table S2, we have calculated the self-shielding factors for the used Pu and Th samples for both employed incident energies $E_i$ = 250 meV and 500 meV, respectively, using Eqs. (S3)-(S6). As can be seen clearly the absorption for the high incident energies use for this experiment, lead to relatively small corrections, in particular for the Th sample.

In order to verify whether the approximation of a flat-plate geometry indeed gives the correct *SSF* we have performed two independent checks. Both tests are based on the reasonable assumption that the self-shielding factor $SSF_{TH}$ for the Th sample is correct due to its weak absorption cross-section. Moreover, the correction factor *C* is expected be reliable because it is independent of the exact sample geometry, and only depends on well-known quantities, such as the sample masses and scattering length. We have carried out both tests for $E_i$ =250 and 500 meV, but in the following illustrate the procedure for $E_i$ =500 meV.

The first test relies on a visual inspection of the data shown in Fig. S2 and the knowledge that because of absence of magnetic scattering at high momentum transfer *Q* due to the magnetic form factor (cf. Fig. 2 in the main text) and the identical structure of δ-Pu and Th, the high momentum transfer scattering that is mostly composed from phonon, multi-phonon, and other multiple scattering processes should be identical. Fig. S2 compares the observed intensities for both the δ-Pu and the Th sample, where panels A and E show the unaltered intensities of both samples. For a meaningful comparison between δ-Pu and Th we subtract the intensity collected for the empty double-wall Al can under the same conditions from both samples according to the first and second part of the sum of Eq.(S1), respectively, employing the corresponding self-shielding factors from Table S2. Moreover, for the Th data the result is also scaled by the correction factor *C* = 0.61 that accounts for different sample mass and scattering cross-section of the two samples. Fig. 2B and F use the self-shielding factor that we calculated for δ-Pu ($SSF_{Pu}$ = 0.92, see table S2) and Th ($SSF_{Th}$ = 0.99), and look relatively similar. There are, however, distinct differences at lower momentum transfer values, which is due to the additional magnetic scattering in δ-Pu. In addition, there are also differences at higher momentum transfers, approximately above *Q* = 7.5 Å. We have introduced a dashed black line in Figs. S2 B-D and F to highlight this small difference. We note that this line represents a somewhat arbitrary choice. It marks the line below which the high-*Q* scattering of Th reaches 3 in arbitrary intensity units. However, we have repeated this procedure for different values of intensity, and the result of this analysis is independent of the value chosen. For δ-Pu this high momentum transfer scattering has lower intensity compared to Th, as can be recognized by the blue color below the dashed black line in Fig. S2B. As explained above this difference is not expected, and as we show in the following is due to the approximation of a flat-plate geometry for the δ-Pu sample, which underestimates $SSF_{Pu}$ because of its strong neutron absorption. The smaller intensity in Fig. 2B (δ-Pu) vs F (Th) highlights that our calculated self-absorption factor for δ-Pu is too small, which means that we subtract a too large fraction of the intensity scattered at the double-wall-Al sample container. In Figs S2C and D we have systematically decreased $SSF_{Pu}$ to 0.88 and 0.83, respectively. We see that this improves the agreement for the high-Q scattering, and for $SSF_{Pu}$ = 0.83 (Fig. S2D) the high-Q scattering for δ-Pu looks identical to the scaled data set for Th (Fig. S2F). Compared to the calculated $SFF_{Pu}$ of 0.92 this represents only a small systematic error of 10% and illustrates that the used flat-plate geometry for the calculation of the SFF still represents a reasonable approximation. An identical procedure for our $E_i$ =250 meV data set shows that in this case $SFF_{Pu} = 0.77$ is the correct self-shielding factor corresponding to a 13% correction from the analytically calculated value shown in Table S2. That the correction is larger for the $E_i = 250$ meV data set compared to $E_i = 500$ meV is consistent, because the absorption cross-sections decrease with increasing incident energy (see Table S2).

The second procedure relies on our knowledge of the analytical shape of the high-$Q$ inelastic scattering that is composed of two contributions, namely a general background that is momentum-transfer-independent, and phonon and multi-phonon scattering that is well-understood. This approach has already been successfully used for various *f*-electron materials (*16*). For a cubic, polycrystalline sample the scattering from phonons is proportional to the scattering function (*39*)

$$S_{ph}(Q,\omega) \propto n(\omega) e^{-2W} \frac{\left\langle (\mathbf{Q} \cdot \varepsilon)^2 \right\rangle}{\omega} G(\omega) \propto Q^2. \tag{S7}$$

Here $n(\omega) = 1/\left(1 - e^{-\hbar\omega/k_B T}\right)$ is the Bose factor, $e^{-2W}$ is the Debye-Waller factor, $G(\omega)$ is the phonon density of states, and the average $\left\langle (\mathbf{Q} \cdot \varepsilon)^2 \right\rangle$ is taken over all phonon modes with the energy $\hbar\omega$. We note that for δ-Pu the phonon cut-off is known to be 12 meV (*26*), and so in principle phonon scattering should be irrelevant for the energy transfers of $\hbar\omega > 50$ meV investigated here. However, at high-Q multi-scattering from other phonons, and Bragg scattering and background scattering modifies this situation. In particular, the δ-Pu sample was encased in a double-wall-Al can and an additional vacuum shield of the sample space also made from Al. Double- or multi- scattering from the sample can or vacuum shield implies an increased flight-path for the scattered neutrons that results in an increased time-of-flight and therefore "virtual" shifts in energy transfer that can easily reach on the order of 50 meV for the incident energy $E_i = 500$ meV. Thus, scattering that is quadratic in the momentum transfer is expected even at much higher energies.

Overall, the scattering at high $Q$ therefore can be modeled with the following expression

$$I(Q,\omega) = A(\omega) + B(\omega) Q^2. \tag{S8}$$

In Figs. S3 A-C we have used this equation to fit the high-$Q$ behavior of $Q$-scans for various energy transfers for the Th sample as denoted by the black solid lines. For direct comparison with the δ-Pu sample the data has been scaled with the correction factor $C$, and the scattering from the Al-can has been subtracted using the calculated self-shielding factor $SSF_{Th} = 0.99$. In Figs. S3 D-F, we show the identical cuts as in A-C, but for the δ-Pu sample. In each panel, the $Q$ scan has been plotted for various values of $SSF_{Pu}$. The black solid line reproduces the best fit of the Th data to Eq. (S8) found in panels A-C. From all $Q$-scans illustrated in Fig. S3, it is apparent that choosing $SSF_{Pu} = 0.83$ for the δ-Pu reproduces the high-$Q$ behavior of the Th sample identified via fits of Eq. (S8). The correct value $SSF_{Pu} = 0.83$ is identical to the one identified independently via the visual method described above, and therefore it will be used for the further analysis of our data presented below. Similarly, this method also reproduces the value for $SSF_{Pu}$ for the $E_i = 250$ meV data set.

As expected at low $Q$, the $Q$-scans for δ-Pu are significantly different from $Q$-scans for Th due to the presence of high-energy magnetic scattering. Already from this data it is clear that the observed additional intensity in δ-Pu roughly follows a magnetic form factor (cf. Fig. 2 in the main text), which unambiguously identifies this contribution as magnetic scattering. We finally want to highlight, while the high-$Q$ intensity is in some cases very sensitive on the exact choice of $SSF_{Pu}$ (see Figs. S3 D and E), this is not true for the low-$Q$ intensity due to the magnetic scattering, which remains qualitatively the same for all used values of $SSF_{Pu}$ with quantitative changes being less than 20%. This illustrates that our chosen method to isolate the magnetic scattering contribution from all other contributions via Eq. (S1) is robust.

**Dynamic Magnetic Susceptibility and Normalization of Data on an Absolute Scale:** Using the correction factor $C$ and the self-shielding factors $SSF_{Pu}$ and $SSF_{Th}$ determined in the previous section, we obtain the purely magnetic contribution to scattering of δ-Pu for both incident neutron energies $E_i$ =500 and 250 meV employed for our experiments. The magnetic contribution to the neutron scattering cross-section is given by (*39*)

$$\frac{d^2\sigma}{d\Omega d\omega} = \frac{N}{\hbar}\frac{k_f}{k_i}\left(\frac{\gamma r_0}{2}\right)^2 e^{-2W} F^2(Q) \sum_{\alpha,\beta}\left(\delta_{\alpha\beta} - \hat{Q}_\alpha \hat{Q}_\beta\right) S^{\alpha\beta}(\mathbf{Q},\omega), \qquad (S9)$$

where $N$ is the total number of unit cells, $(\gamma r_0/2) = 0.2695 \times 10^{-12}$ cm is the magnetic scattering length, $k_i$ and $k_f$ are incident and final scattering vector. $F(Q)$ is the magnetic form factor. $\hat{Q}$ is a unit vector parallel to the scattering vector, and $\alpha, \beta$ denote Cartesian coordinates $x$, $y$ and $z$. $S^{\alpha\beta}(Q,\omega)$ is the dynamic spin-correlation function that is related to the imaginary part of dynamic magnetic susceptibility $\chi''(\mathbf{Q},\omega)$ via the fluctuation-dissipation theorem (*40*)

$$\chi^{\alpha\beta}{}''(\mathbf{Q},\omega) = g^2 \mu_B^2 \frac{\pi}{\hbar}\left(1 - e^{-\hbar\omega/k_B T}\right) S^{\alpha\beta}(\mathbf{Q},\omega). \qquad (S10)$$

As demonstrated in Ref. (*41*), assuming that the instrumental resolution is "decoupled" from the scattering response, $S^{\alpha\beta}(Q,\omega)$ is related to the measured intensity by

$$I(\mathbf{Q},\omega) = \frac{N}{\hbar}\left(\frac{\gamma r_0}{2}\right)^2 e^{-2W} F^2(Q) \tilde{S}(\mathbf{Q},\omega) k_f R_0, \qquad (S11)$$

where $\tilde{S}(\mathbf{Q},\omega) = \left(\delta_{\alpha\beta} - \hat{Q}_\alpha \hat{Q}_\beta\right) S^{\alpha\beta}(\mathbf{Q},\omega)$, and $R_0$ is the instrument-dependent resolution volume. Here we have determined $R_0$ by measuring a standard vanadium sample for the identical instrument setup used for all our measurements as described in Ref.(*41*). Further, for the case of isotropic spin-fluctuations as in a paramagnetic phase such as δ-Pu $S^{xx}(Q,\omega) = S^{yy}(Q,\omega) = S^{zz}(Q,\omega)$, and thus $\tilde{S}(Q,\omega) = 2S^{zz}(Q,\omega)$. Finally, for spin-fluctuations driven by a Kondo impurity the dynamic spin-correlation function has no $Q$-dependence with $\tilde{S}(\mathbf{Q},\omega) = \tilde{S}(\omega)$ and the only $Q$-dependence of the observed intensity is given by the magnetic form factor $F(Q)$. Using Eqs. (S9)-(S11), we then obtain the imaginary part of the dynamic magnetic susceptibility in absolute units as

$$F^2(Q)\chi''(\omega) = F^2(Q)\chi^{zz}{}''(\omega) = g^2\mu_B^2 \frac{\pi}{2}\left(1 - e^{-\hbar\omega/k_B T}\right)\frac{13.77(\text{barn}^{-1})I(Q,\omega)}{e^{-2W}Nk_f R_0}. \qquad (S12)$$

The result of this normalization is shown in Fig. S4. The dynamic magnetic susceptibilities obtained for $E_i$ =500 (Fig. S4A) and 250 meV (Fig. S4B), respectively, look identical apart from differences in energy resolution, and a difference in the absolute scale of approximately 30%. We note that this is expected because 30% is roughly the accuracy of this standard way of normalizing magnetic scattering on an absolute scale. In Fig. S4C we also show energy scans obtained by integrating the data from $Q = 2$ to $6$ Å$^{-1}$ for both incident energies. The intensity at high energies above energy transfers of approximately 120 meV the intensity for the $E_i$ =250 meV data seems too large when compared to the 500 meV data set. This is likely due to the limited $Q$-range of the $E_i$ =250 meV data set at higher energy transfers, which typically leads to normalization problems for Q-integrated data sets.

**Sum-Rules for Neutron Scattering:** The imaginary part of the dynamic magnetic susceptibility $\chi''(\omega)$ is related to both the fluctuating magnetic moment, as well as the static susceptibility via so-called sum-rules. The square of the fluctuating magnetic moment can be obtained by integrating over $\chi''(\omega)$ via

$$\langle \mu_z^2 \rangle = \frac{1}{\pi} \int_{-\infty}^{\infty} d\omega \left[ n(\omega) \chi''(\omega) \right], \tag{S13}$$

where $n(\omega)$ is the Bose factor as before. We note that usually, to obtain the correct value for the moment, the dynamic magnetic susceptibility in Eq. (S13) has also to be integrated in momentum-space over an entire Brillouin zone, and then normalized to the size of the Brillouin zone. Here this is not required because we assume that $\chi''(\mathbf{Q},\omega) = \chi''(\omega)$.

The static magnetic susceptibility as measured via bulk measurements can be obtained via the following sum-rule

$$\chi'(0) = \frac{1}{\pi} \int d\omega \frac{1}{\omega} \chi''(\omega), \tag{S14}$$

as also described in the main text. We note that strictly speaking because the dynamic magnetic susceptibility gnerally depends on both the momentum and energy transfer, and is thus given by $\chi''(Q,\omega)$, this sum-rule usually gives the static susceptibility $\chi'(Q,0)$ at the momentum transfer $\mathbf{Q}$ measured by the neutron scattering experiment. Frequently $\chi'(\mathbf{Q},0)$ is proportional to $\chi'(0)$, but magnetic susceptibility measurements do not measure $\chi'(\mathbf{Q},0)$. Here, however, we assume that the dynamic magnetic susceptibility is independent of $\mathbf{Q}$ and given by $\chi''(\omega)$.

To calculate $\langle \mu_z \rangle$ and $\chi'(0)$ via Eq. (S13) and (S14), we have parametrized $\chi''(\omega)$ by fitting Eq. **(1)** in the main text to the two data sets shown in Fig. S4C. Here a damped Lorentzian function was employed for each of the peaks in the data. The results of this analysis for both employed incident energies and their average are provided in Table S3.

**Other Possible Sources for the Observed Magnetic Scattering:** We note that both crystal field excitations or intermultiplet transitions would, in principle, lead to similar forms of the dynamic susceptibility $\chi''(\omega)$ observed for δ-Pu. However, the former will be generally too broad in an actinide to be observable by neutron scattering, especially at room temperature. Moreover, the effects of crystal field splittings are included in the DMFT theory used to calculate the dynamic magnetic susceptibility shown in Fig. 1F in the main text. Intermultiplet transitions are known to exist at only much higher energies for Pu ions. The value for the first excited state in a $5f^5$ ionic system is ~ 400 meV (*42*). Although this could be lowered (and broadened) slightly by screening from the conductions electrons (*43*), the value is still likely to be above 300 meV. Such intermultiplet scattering will also have a different form factor to that observed here (*43*).

**Contributions of the 5$f^6$ Electronic Configuration to the Magnetic Form Factor:** Because the spatial extent of the 5$f$ electrons is non-zero compared with the neutron wavelength of the scattered neutrons, the magnetic scattering is not independent of the momentum transfer $Q$ and is described via the magnetic form factor $F(\mathbf{Q}) = \int d\mathbf{Q}(\mathbf{M}(\mathbf{r})e^{-i\mathbf{Q}\cdot\mathbf{r}})$ that represents the Fourier transform of the magnetization density $\mathbf{M}(\mathbf{r})$ (*39*). $F(\mathbf{Q})$ can be separated into the individual spin ($\mu_S = \int d\mathbf{r}\,\mathbf{M}_S(\mathbf{r})$) and orbital ($\mu_L = \int d\mathbf{r}\,\mathbf{M}_L(\mathbf{r})$) components to the magnetization density, and the total magnetic amplitude may then be written as $\mu f(\mathbf{Q}) = \mu_S f_S(\mathbf{Q}) + \mu_L f_L(\mathbf{Q})$. In the dipole approximation this can be rewritten as (*21*):

$$\mu f(\mathbf{Q}) = \mu_S \langle j_0 \rangle + \mu_L (\langle j_0 \rangle + \langle j_2 \rangle). \tag{S15}$$

Here the $\langle j_l \rangle$ are defined via $\langle j_l(Q) \rangle = \int_0^\infty dr \left[ U^2(r) j_l(Qr) \right]$, where $U^2(r)$ is the probability distribution of the radial part of the 5$f$ single-electron wave function, and the $j_l(Qr)$ is the spherical Bessel function. The $\langle j_l \rangle$ are tabulated for the various 5$f$ configuration (*20*), but in general do not change significantly for different 5$f$ configurations.

By definition all form factors at $Q = 0$ are equal to unity, and one obtains $\mu = \mu_S + \mu_L$. Because the 5$f^6$ configuration is nonmagnetic, we have $\mu = \mu_S + \mu_L = 0$, and therefore $\mu_S = -\mu_L$. This implies that 5$f^6$ state does not contribute to the magnetic scattering at $Q = 0$, but only at non-zero $Q$. Using Eq. (S15), we see that the magnetic amplitude for the 5$f^6$ configuration at $Q \neq 0$ is described via $\mu f(\mathbf{Q}) = \mu_L \langle j_2 \rangle$. We do not know $\mu_L$ for the 5$f^6$ state but it is likely to be similar to the value for the 5$f^5$ state. Recalling that at $Q = 0$, $\langle j_0 \rangle = 1$ and $\langle j_2 \rangle = 0$, as well as that the maximum contribution of $\langle j_2 \rangle \approx 0.2$ at $Q = 3$ Å$^{-1}$ (Q/4π=0.24 Å$^{-1}$, cf. Fig. 2 in the main text), we can estimate that the contribution of the 5$f^6$ configuration to the magnetic form factor for $Q \neq 0$ is at least a factor five smaller compared to the 5$f^5$ configuration. In addition, as shown in Table 1 in the main text, the occupation of the 5$f^6$ state may be up to a factor three smaller than for the 5$f^5$ state. Because the neutron intensity is proportional to the square of the magnetic form factor, the sensitivity of our experiment to the contribution of the 5$f^6$ state to the magnetic form factor is only 0.4-4% of the 5$f^5$ state.

**Details of the Dynamical Mean Field Theory calculation:** In the following we describe a few details and crosschecks of our DMFT calculations. To crosscheck the validity of our analytic continuation, we also calculated the fluctuating magnetic moment $\mu(E)$, which is shown in Fig. S6. The fluctuating moment up to the energy $E$ is defined by the integral of the dynamic magnetic susceptibility (cf. Eq. (S13)) by the following formula

$$\langle \mu^2(E) \rangle = \frac{1}{\pi} \int_{-\infty}^{E} d\omega \left[ (1 + n(\omega)) \chi''(\omega) \right]. \tag{S16}$$

The value at $E = \infty$ can also be expressed by the equal time correlation function $\mu^2 = \langle \vec{\mu} \cdot \vec{\mu} \rangle$. The latter is directly sampled by the CTQMC method, and can be compared with the integral of the analytically continued susceptibility. It turns out that the two are equal to within $0.001\mu_B$.

The magnetic form factor $F(Q)$ is the Fourier transform of the spatial distribution of the electronic magnetic moment, i.e.,

$$F(Q) = \frac{-1}{2\mu_B} \langle M_\perp(Q) \rangle, \quad (S17)$$

where $M_\perp(Q)$ is the component of the magnetization density that is perpendicular to the scattering vector $Q$ in the neutron scattering experiment. The formalism to compute the form factor within the DFT+DMFT framework was developed in Ref.(*44*), with more tests and details given in Ref. (*45*).

To estimate the Brillouin-zone-averaged dynamic spin-correlation function (cf. Eq.(S9)), which is measured in a polycrystalline sample, we use

$$S(Q,\omega) = F^2(Q) \frac{1/2}{1 - e^{-\beta\hbar\omega}} \chi''(\omega). \quad (S18)$$

Alternatively, we have also calculated the local dynamical magnetic susceptibility defined by a joint function of *f*-electron spectral density as obtained from LDA+DMFT (CTQMC)(*33*). The result from this calculation (not shown here) agrees reasonably well with our experiment and the dynamical susceptibility calculated via Eq. (4) in the main text. Notice that the result from Eq. (4) should be more rigorous because it contains the vertex correction.

An exact property of the dynamic magnetic susceptibility is that $\chi''(\omega) = -\chi''(\omega)$, which implies that $\chi''(\omega = 0) = 0$. As shown in Fig. S6, the dynamical magnetic susceptibility calculated for δ-Pu as described above via Eq. (4) in the main text and (S18) fulfills this condition.

Finally, we discuss the small discrepancies of the dynamical magnetic susceptibility $\chi''(\omega)$ obtained via our LDA+DMFT calculations with respect to our experimental results as shown in Fig. 1D in the main text. The position of the main peak, the linear fall off at low frequencies and the broad distribution of the spectral weight extending to high energies, are all very robust features of both the LDA+DMFT calculations and the experimental data. While the LDA+DMFT calculation does not generate a second distinct feature at higher energy, and for energies below approximately 70 meV it exhibits more spectral weight than the experimentally determined $\chi''(\omega)$, taking into account the respective uncertainties of experiment and theory shows that they are in quantitative agreement. The uncertainties in the theory are due to the analytic continuation from Matsubara frequency to real frequencies carried out by the maximum entropy method and are more pronounced at intermediate energies. The experimental uncertainties are larger at very low energies and at high energies. The experimental data at low energies exhibits larger uncertainties because the subtraction of the structural Bragg peaks and phonons from our δ-Pu data via the non-magnetic Th analogue is less reliable at low energy transfers due to the slightly different lattice constants of δ-Pu and Th. However, for energy transfers that are larger than our energy resolution (≈40 meV) and above the phonon cut-off (≈15 meV, see Ref. ( *26*)) the method works reliably. At high energies the uncertainties are larger simply due to lower neutron statistics that are a result of the kinematic scattering conditions for neutrons that restrict the accessible range in momentum transfers *Q* with increasing energy transfer, in particular for low *Q*. The limited *Q*-range can lead to artificially increased intensities for higher energies. We note that this is consistent with the fact that the high-energy feature in $\chi''(\omega)$ is more pronounced for the data set collected with an incident energy $E_i$ = 250 meV, for which the $Q$-range at high energies is more limited (cf. Fig. S4 A and B).

**References and Notes**

See references 35 to 44 of main text

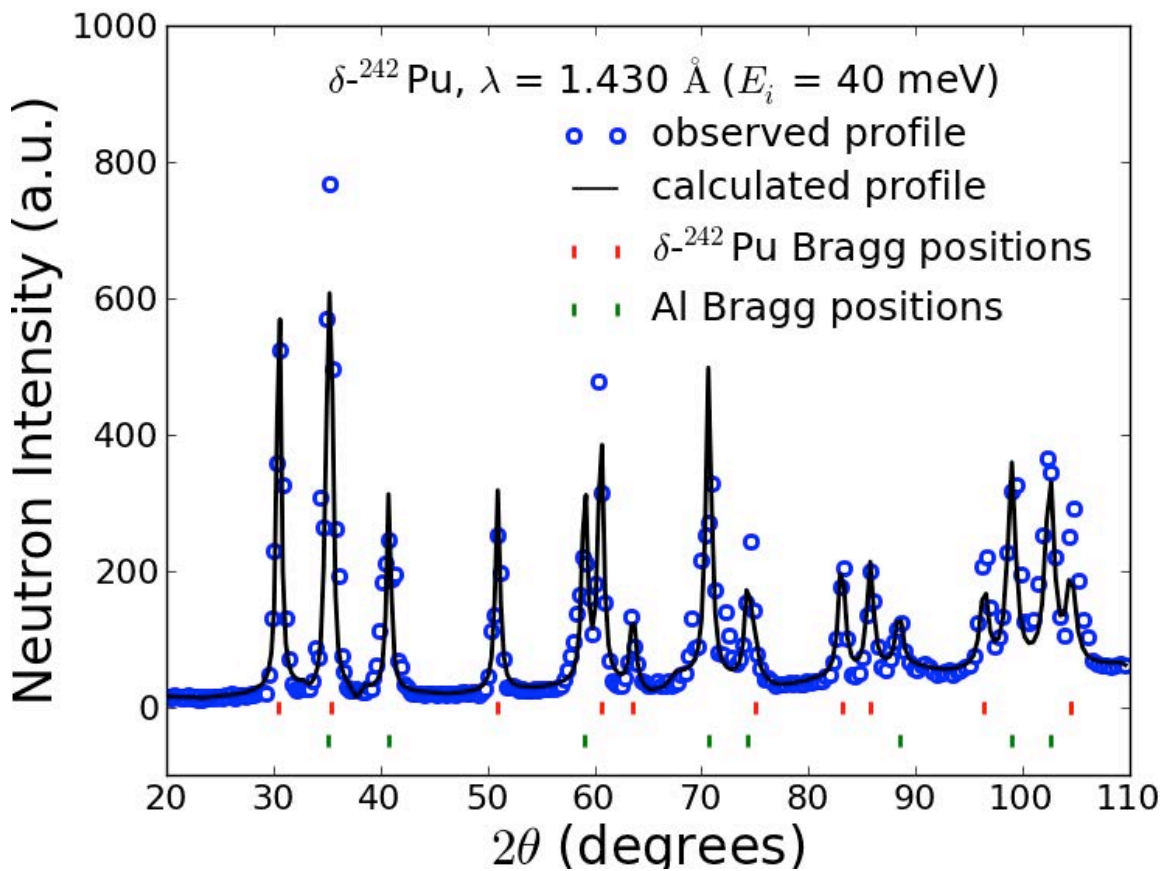

**Fig. S1.** A neutron powder diffraction profile of the δ-Pu sample enclosed in a double-wall-Al can used for the experiments reported here is shown. The data were collected on ARCS at room temperature using an incident energy $E_i$ = 40 meV (λ = 1.43 Å), and the obtained spectrum was integrated over all recorded energy transfers. Blue empty circles denote the data, the black solid line is a Rietveld fit, and the red and green vertical bars denote the positions of the δ-Pu and Al Bragg reflections, respectively, where the latter stem from scattering at the double-wall-Al sample container.

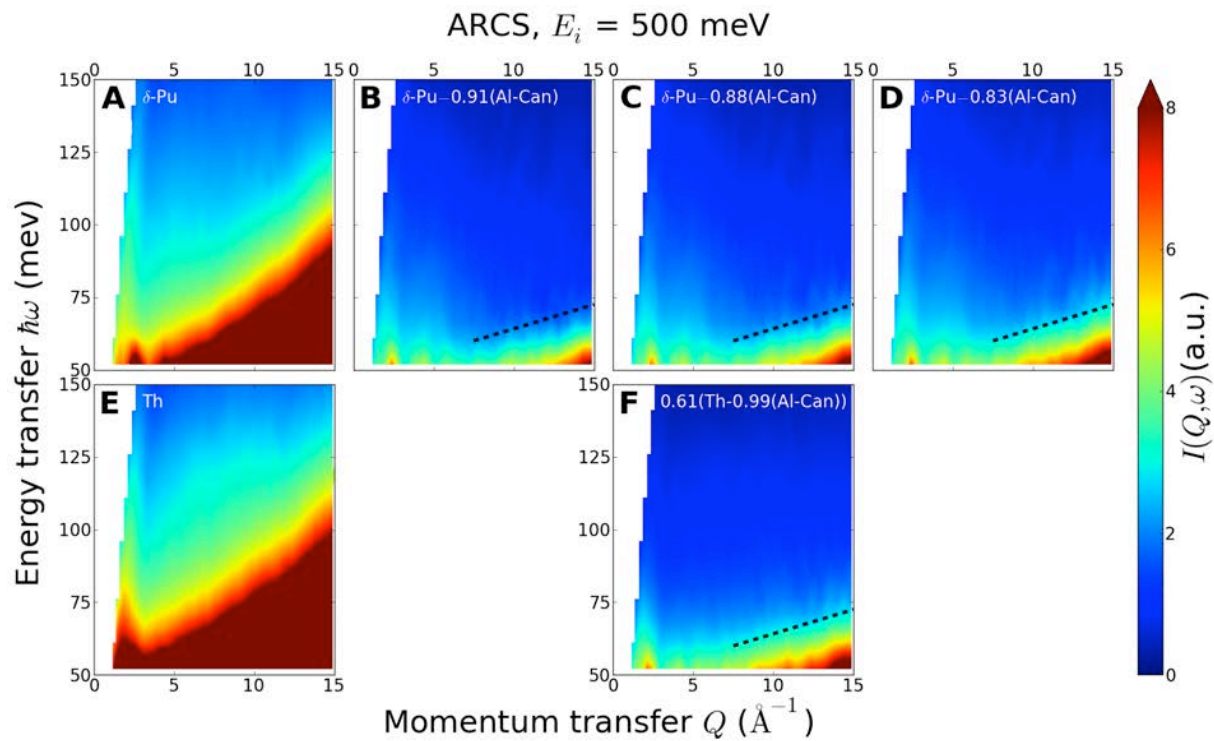

**Fig. S2.** Visual method of adjusting the analytically calculated self-shielding factor $SSF_{Pu}$ for δ-Pu. The intensities for both the δ-Pu (top row) and the Th (bottom row) sample at room temperature as measured on ARCS with $E_i$ = 500 meV and identical statistics are compared. Panels (**A**) and (**E**) show the unaltered intensities observed for δ-Pu and Th, respectively. Panel (**B**) shows the intensity for δ-Pu, however, with the intensity of empty double-wall-Al can subtracted, where the calculated self-shielding factor $SSF_{Pu}$ = 0.92 has been employed (compare Table S2 and Eq. (S1)). In (**C**) and (**D**) the same same subtraction is illustrated, however, with $SSF_{Pu}$ set to 0.88 and 0.82, respectively. Similarly, in (**F**) the intensity of the double-wall-Al can has been subtracted from the Th data using the calculated self-shielding factor $SSF_{Th}$ = 0.99, where additionally the intensities have been scaled by the correction factor $C$ = 0.61 for direct comparison with panels (**B**)-(**D**) for δ-Pu. For the details of the comparison, and particular the meaning of the dashed black line in panelds (**B**)-(**D**) and (**F**), please see the main text.

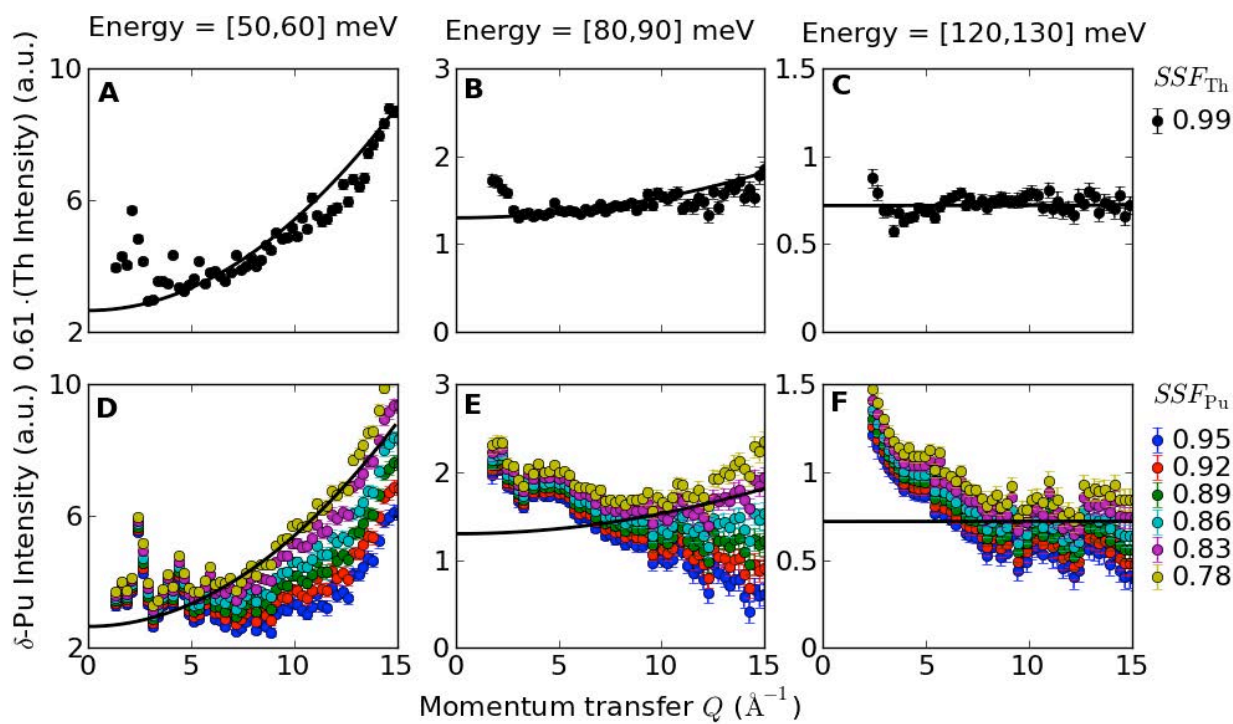

**Fig. S3.** Analytical method of adjusting the analytically calculated self-shielding factor $SSF_{Pu}$ for δ-Pu. (**A**)-(**C**) Momentum transfer $Q$-scans for the Th sample that were obtained by energy integrating the data over the energy range given on top of the panels. The data were scaled by the correction factor $C=0.61$ (see text) and the scattering from the double-wall-Al can was subtracted according to Eq. (S1) with a self-shielding factor $SSF_{Th} = 0.99$. The solid black line in each panel is a fit to Eq. (S8) that describes the high-$Q$ dependence of the data. (**D**)-(**F**) the identical $Q$-scans are shown for the δ-Pu where the scattering from the double-wall-Al can was subtracted using various values of $SSF_{Pu}$. The black solid line reproduces the fit that was made to the Th data in (**A**)-(**C**).

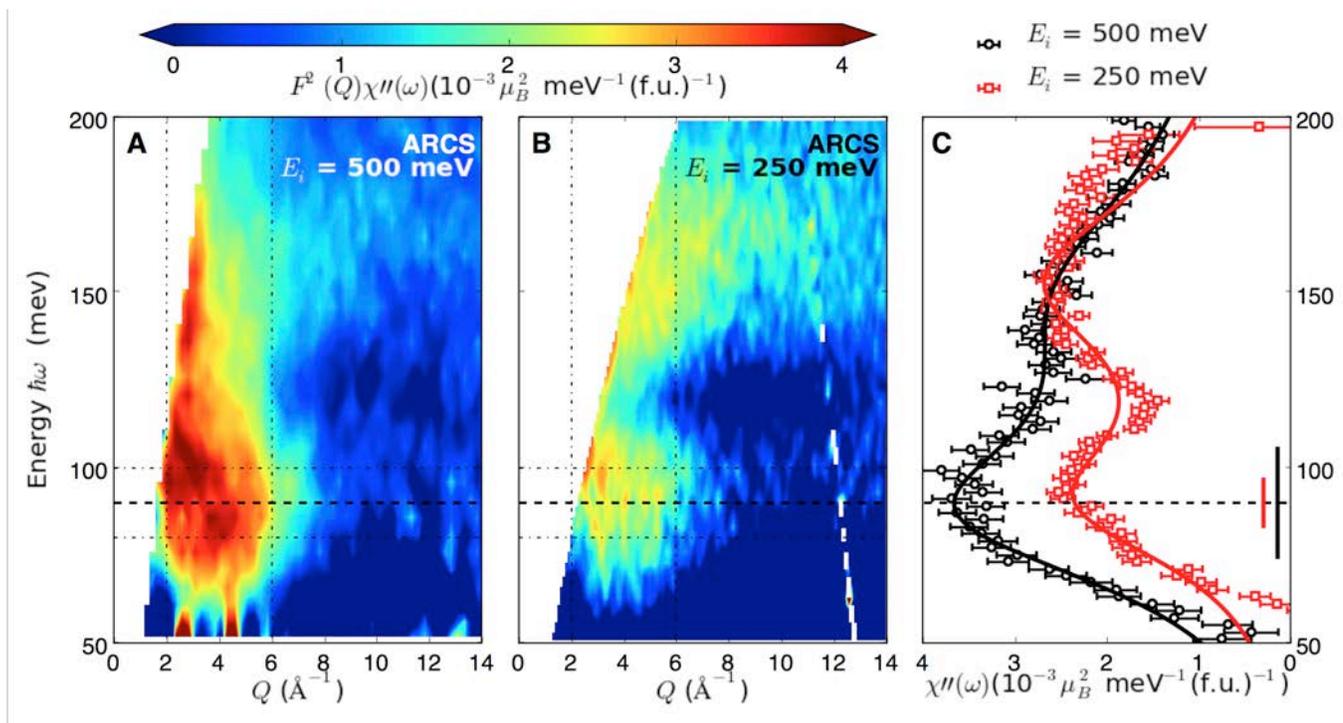

**Fig. S4.** The dynamic magnetic susceptibility of δ-Pu measured via neutron spectroscopy for incident neutron energies of $E_i$ =500 (**A**) and 250 meV (**B**), respectively. In (**C**) energy-cuts obtained by integrating over the momentum transfer $Q$-ranges denoted by the vertical dash-dotted black lines in (**A**) and (**B**) are shown. The black and red solid lines denote fits to Eq. (1) in the main text for $E_i$ =500 and 250 meV, respectively. The black and red vertical bars represent the corresponding energy resolutions.

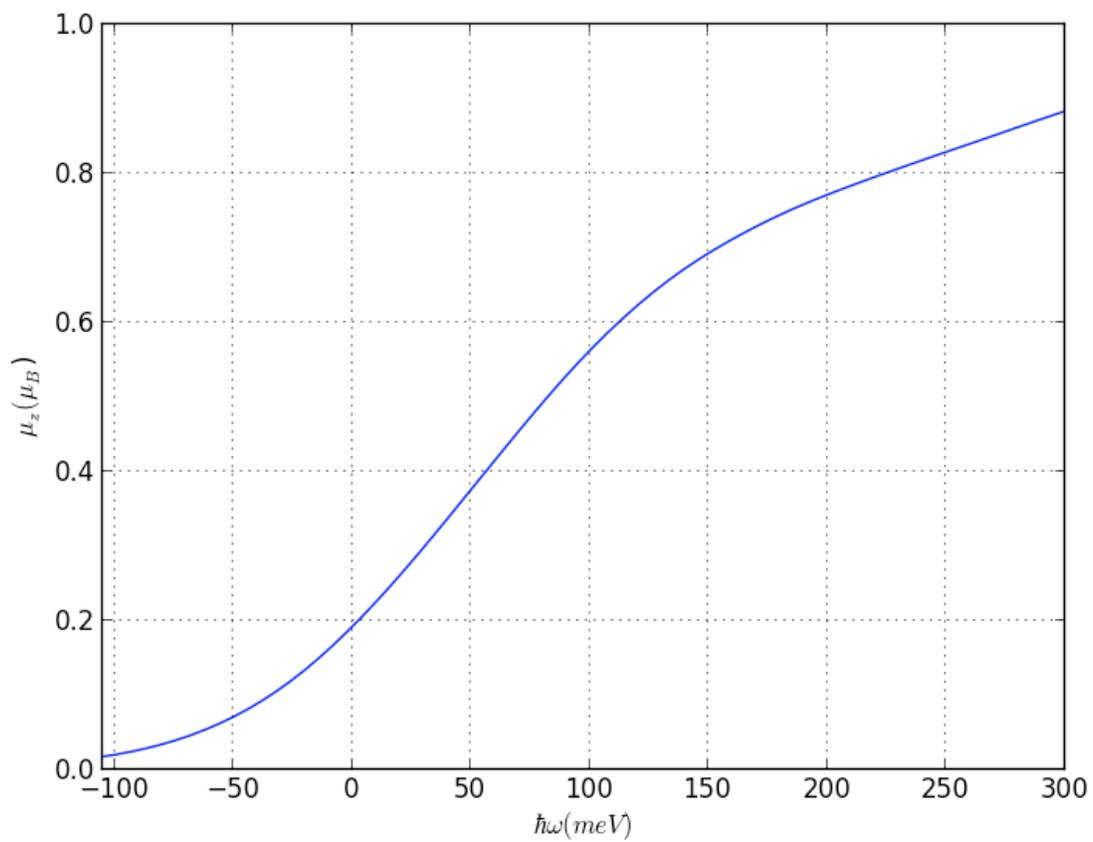

**Fig. S5**: Theoretical fluctuating magnetic moment of δ-Pu defined in Eq.(S16).

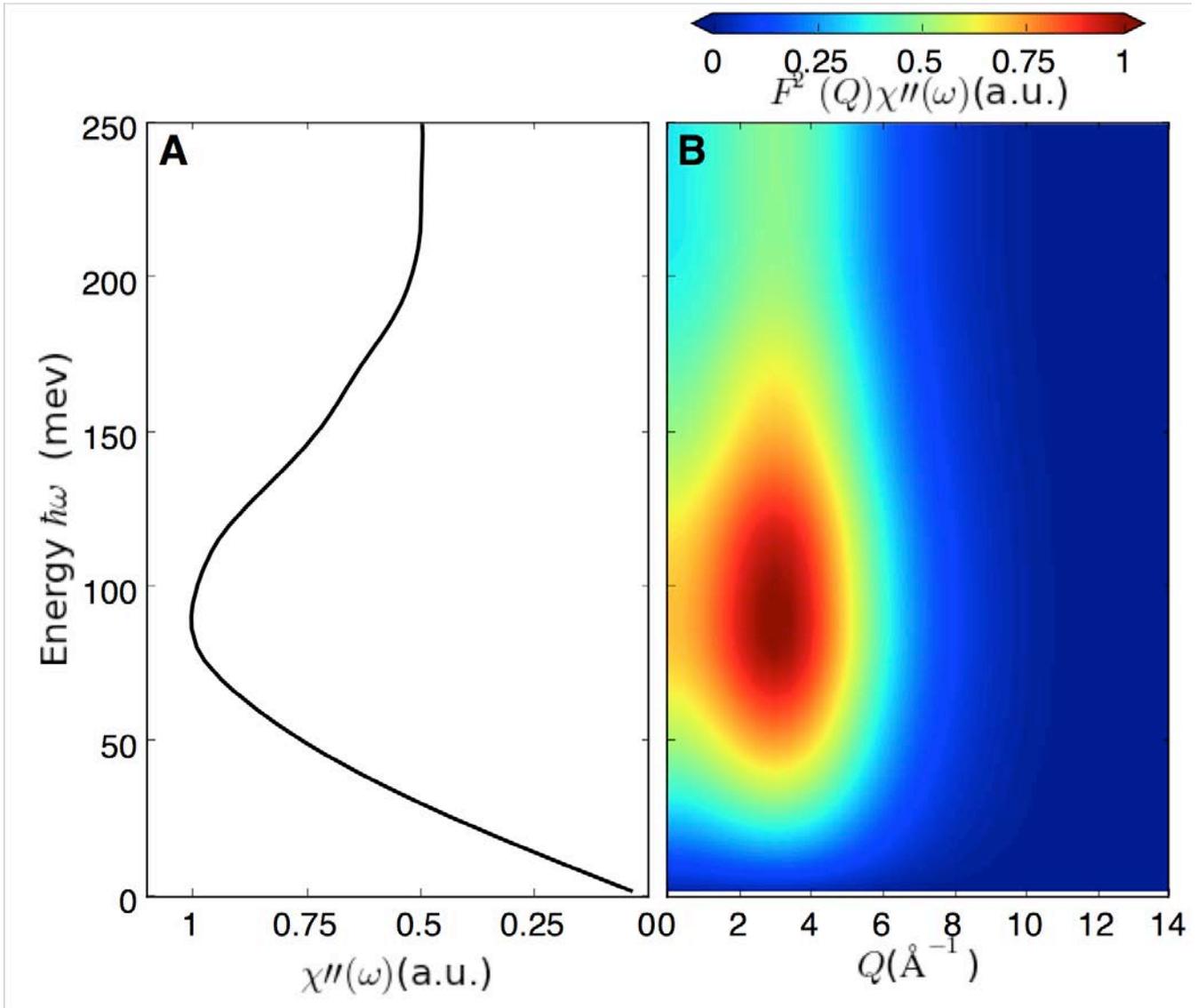

**Fig. S6**: Summary of the main results of our DMFT calculation. (**A**) The dynamic magnetic susceptibility $\chi''(\omega)$ of δ-Pu as calculated via Eq. (4). (**B**) The full momentum transfer ($Q$) and energy transfer ($\hbar\omega$) dependence of the calculated magnetic fluctuation spectrum $F^2(Q)\chi''(\omega)$ for δ-Pu is shown. The $Q$-dependence originates (see text) from the magnetic from factor $F(Q)$ that was calculated via Eq. (S17).

**Table S1.** Isotopics and coherent and absorption neutron cross-sections of the δ-Pu sample used for the experiment described in this report. We note that the tabulated absorption cross-section is for thermal neutrons with an energy of 25 meV.

| Isotope | Weight (%) | $\sigma_{abs}$ (barn) | $\sigma_{coh}$ (barn) | Molar weight (g/mol) |
|---|---|---|---|---|
| $^{238}$Pu | 0.6426 | 558(7) | 25.0(1.8) | 238.050 |
| $^{239}$Pu | 1.4937 | 1017.3(2.1) | 7.5 | 239.052 |
| $^{240}$Pu | 4.6938 | 289.6(1.4) | 1.54 | 240.054 |
| $^{241}$Pu | 0.5789 | 1375 | 9.0 [10] | 241.057 |
| $^{242}$Pu | 92.59 | 18.5(5) | 8.2 | 242.059 |
| Total | 100 | 57.54 | 7.989 | 241.886 |

**Table S2.** Self-shielding factors SSF for the δ-Pu and Th samples for both employed incident energies $E_i$ = 250 meV and 500 meV obtained via Eqs. (S3)-(S6). The densities used in Eq. (S4) were $\rho_{\delta-Pu}$ = 15.81 g/cm$^3$ and $\rho_{Th}$ = 12.7 g/cm$^3$, respectively.

| Sample | $E_i$ (meV) | $\Sigma_s$ (cm$^{-1}$) | $\Sigma_a$ (cm$^{-1}$) | $\Sigma_t$ (cm$^{-1}$) | SSF$_0$ | SSF |
|---|---|---|---|---|---|---|
| δ-Pu | 25.3 | 0.317 | 2.281 | 2.598 | 0.695 | 0.721 |
| | 40.0 | | 2.047 | 2.364 | 0.716 | 0.774 |
| | 250.0 | | 0.819 | 1.136 | 0.847 | 0.885 |
| | 500.0 | | 0.579 | 0.896 | 0.877 | 0.918 |
| Th | 25.3 | 0.406 | 0.224 | 0.626 | 0.912 | 0.967 |
| | 250.0 | | 0.080 | 0.486 | 0.931 | 0.988 |
| | 500.0 | | 0.057 | 0.463 | 0.934 | 0.991 |

**Table S3.** Results of the sum-rule analysis for both the fluctuating magnetic moment $\langle \mu_z \rangle$ (Eq. (S13)) and the static susceptibility $\chi'(0)$ (Eq. (S14)) are provided for incident energies $E_i$ = 250 meV and 500 meV. The error bars provided in brackets are statistical in nature and stem from the statistical errors associated with counting neutrons. We note that the systematic error associated with the process of normalizing the neutron counts on an absolute scale is much larger and about 30% (see text). Therefore, we have used the systematic error for these quantities in the main text.

| Incident Energy $E_i$ (meV) | $\langle \mu_z \rangle$ ($\mu_B$) | $\chi'(0)$ ($10^{-4}$ cm$^3$/mol) |
|---|---|---|
| **250** | 0.53(2) | 0.63(2) |
| **500** | 0.69(1) | 0.97(2) |
| **Average** | 0.6(1) | 0.80 (2) |